\documentclass[epj]{svjour}
\usepackage{epsfig}
\usepackage{amsmath}
\newcommand{\beq}{\begin{eqnarray}}
\newcommand{\eeq}{\end{eqnarray}}
\newcommand{\be}{\begin{eqnarray*}}
\newcommand{\ee}{\end{eqnarray*}}

\newcommand{\cc}{{c \bar{c}}}

\begin{document}
\title{Can the RHIC $J/\psi$ puzzle(s) be settled at LHC?}
\author{L. Bravina\inst{1} \and A. Capella\inst{2} \and E.G. Ferreiro\inst{3} 
\and A.B. Kaidalov\inst{4} \and K. Tywoniuk\inst{1} \and E. Zabrodin\inst{1}} 

\institute{
  Department of Physics, University of Oslo\\
  0318 Oslo, Norway
  \and
  Laboratoire de Physique Th\'eorique\thanks{Unit\'e Mixte de
    Recherche UMR n$^{\circ}$ 8627 - CNRS}, Universit\'e de Paris XI,
  B\^atiment 210, \\
  91405 Orsay Cedex, France
  \and
  Departamento de F{\'\i}sica de Part{\'\i}culas, Universidad de
  Santiago de Compostela, \\
  15782 Santiago de Compostela, Spain
  \and
  Institute of Theoretical and Experimental Physics\\
  RU-117259 Moscow, Russia
}
\mail{konrad.tywoniuk@fys.uio.no}

\date{Received: date / Revised version: date}
%
\abstract{
  One observes strong suppression effects for hard probes, e.g. the 
  production 
  of $J/\psi$ or high-$p_T$ particles, in nucleus-nucleus (AA) collisions 
  at RHIC. 
  Surprisingly, the magnitude of the suppression is quite similar to that at 
  SPS. In order to establish whether these features arise due to the 
  presence of a thermalized system of quarks and gluons formed in the course of
  the collision, one should investigate the impact of suppression mechanisms 
  which do not explicitly involve such a state. We calculate shadowing for 
  gluons 
  in the Glauber-Gribov theory and propose a model invoking a 
  rapidity-dependent absorptive mechanism motivated by energy-momentum 
  conservation effects.
  Furthermore, final state suppression due to interaction with co-moving matter 
  (hadronic or pre-hadronic) has been shown to describe data at SPS. We extend 
  this model by including the backward reaction channel, i.e. recombination of 
  open charm, which is estimated directly from pp data at RHIC.
  Strong suppression of charmonium both in pA and AA collisions at LHC is predicted. 
  This is in stark contrast with the predictions of models assuming QGP 
  formation and thermalization of heavy quarks.
  \PACS{
    {13.85.-t}{Hadron-induced high- and super-high-energy
      interactions (energy $>$ 10 GeV)}       \and
    {25.75.-q}{Relativistic heavy-ion collisions} \and
    {25.75.Cj}{Photon, lepton, and heavy quark production in heavy ion
    collisions}
  }
}

\maketitle

\section{Introduction}
Charmonium production off nuclei is one of the most pro-mising probes for studying
properties of matter created in ultrarelativistic heavy ion collisions. Being a heavy 
particle, it can be used as a probe of the properties of the medium created in 
these collisions, such as the 
intensity of interactions and possible thermalization.

The RHIC era in charmonium physics has brought to light further aspects beyond what 
was known during the SPS experiments. The gold plated signal of charmonium suppression
due to colour screening in a quark-gluon plasma was shown to be a characteristic
of models which did not assume such a state and recently
the study of charmonium regeneration, or recombination, has become a field of active
study. In this sense the results from RHIC establish an important step, bridging the 
vast energy jump to the future LHC experiments.

The RHIC results on the centrality-dependent nuclear modification factor of charmonium
in Au+Au collisions at $\sqrt{s} = 200$ GeV \cite{Adare:2007} contained two puzzling 
features. The first was the fact, that the suppression at mid-rapidity coincided
with the level of suppression in 
Pb+Pb collisions at $\sqrt{s} = 17.3$ GeV for the same number of participants
\cite{Alessandro:2004ap}. Since the medium produced at RHIC is denser and lives 
longer than at SPS one would, a priori, expect a stronger suppression with increasing 
energy. 
The second puzzling feature was the stronger suppression at forward rapidity 
than at $y = 0$. Once again, the medium in the central part of the collision is 
the most dense and from simple arguments we would expect an opposite behaviour than
what is seen in the data.
Additionally, data on charmonium in d+Au collisions at the same energy
revealed a decrease of the suppression due to cold nuclear matter compared to 
lower energies.

These seemingly puzzling features in the data bring to light two new aspects of 
charmonium 
physics in nuclear reactions, namely the role of initial state effects and the role
of possible secondary $J/\psi$ production from recombination of open charm. 
At RHIC these effects are visible and important for the detailed description,
at LHC their magnitude, especially the gluon shadowing, will be much larger. 
We argue that only the full description of charmonium dynamics from SPS to LHC energies
will shed new light on the underlying physics of dense partonic systems.

\section{Baseline: initial state effects}
In order to quantify the modifications of charmonium due to the presence of a dense 
partonic medium, one usually starts from the simpler proton-nucleus ($pA$) collisions. 
At least two modifications in the $pA$ case compared to $pp$ are known to come 
into play. On the one hand, charmonium suppression is seen to scale with L, the traversed
path length by the initially formed $\cc$ pair within the nuclear medium. In the 
Glauber model, this leads to a suppression factor $\propto \exp \left( - \rho L 
\sigma_{abs}\right)$, controlled by the so-called absorptive cross section, 
$\sigma_{abs}$. In p+Pb collisions at $\sqrt{s} = $ 19 GeV this cross section was
found to be $\sim 4.5$ mb \cite{Ramello:2003}. 
On the other hand, the nuclear gluon distribution is modified at
high energies, leading to
\beq
R_g^A \equiv \frac{g^A(x,Q^2) }{A g^p(x,Q^2)} \; < \; 1 \;,
\eeq 
at values of Bjorken-x $ < 0.1$, called gluon shadowing. Usually one assumes a
gluon fusion mechanism for the charmonium production, and therefore this effect
is of crucial importance. 

Recently, it has been realized that the
nuclear suppression of charmonium in $pA$ collision exhibits a non-trivial energy 
dependence (see e.g. \cite{Cortese:2008,Woehri:2008}) namely, that $\sigma_{abs}$ is 
decreasing
with energy , $17.3 < \sqrt{s} < 45$ GeV,
in contrast to elementary theoretical expectations \cite{Bedjidian:2004gd}.
At RHIC, the situation is more complex due to the presence of additional gluon shadowing. 
The absorptive cross section introduced
above is replaced by a rapidity-independent $\sigma_{break-up}$ which takes values 
in the range $\{ 0.7, 4.5 \}$ mb, depending on which shadowing parameterization
is chosen \cite{Adler:2005ph,Adare:2007gn}. This approach seems to be in conflict
with the strong $x_F$-dependence of $J/\psi$ suppression observed in 
\cite{Alde:1990wa,Leitch:1999ea}, although the accessible range of $x_F$ at RHIC
is quite  limited.
In summary, there are still many open questions
which seems to lack an explaination within a unified framework.

The observed trends are, in fact, in line with expectations from the 
Glauber-Gribov theory \cite{Boreskov:1992ur,Braun:1997qw}. In this approach 
the quantity 
$\sigma_{abs}$ controls the contribution
of a certain set of diagrams valid at low energies, which corresponds to the 
Glauber model. 
More involved diagrams become, of course, important at high energies \cite{Gribov}.
They correspond
to a space-time picture where the initially small projectile evolves into a large 
fluctuation which can interact coherently with all of the constituents of the target
nucleus. This leads to nuclear shadowing. The transition from the former, planar,
 to the latter, non-planar, regime is governed by a critical energy scale
\beq
\label{eq:sM}
s_M \;=\; \frac{M^2_\cc}{x_+} \, \frac{R_A m_N}{\sqrt{3}} \;,
\eeq
where $M_\cc$ is the mass and $x_+ = (\sqrt{x_F + 4 M_\cc^2/s}
+ x_F)/2$ is the longitudinal momentum fraction of the heavy system.
This transition signals the breakdown of the semiclassical probabilistic picture
of longitudinally ordered multiple scattering as one goes to high energies, $s > s_M$.
In this scenario, the diagrams that are controlled by $\sigma_{abs}$ are cancelled 
in the summation and, instead, shadowing of nuclear partons appear \cite{Capella:2006mb}.

At finite energies there are also well know corrections due to conservation of 
energy-momentum \cite{Capella:1976ef,Boreskov:1992ur}. These are, of course, most
prominent when $x_F \rightarrow 1$, but can also lead to suppression at mid-rapidity.

These features lead us to propose the following model for charmonium suppression
at all energies, see \cite{Arsene:2007gx} for the details. The suppression of charmonium
in pA compared to $A$ elementary pp collisions is given by
\beq
R^\cc_{pA}(x_F,s) = \frac{1}{A} \int d^2b \, S^{FE}(x_+,b) S^{shad}(x_2,Q^2,b) \;.
\eeq 
The finite energy (FE) suppression factor, $S^{FE}(x_+,b)$, is given by \cite{Braun:1997qw}
\beq
\label{eq:Sfe}
S^{FE}(x_+,b) = \left\{ \begin{array}{cl}
  \frac{1}{\tilde{\sigma}} \left[1 - \exp(-\tilde{\sigma} T_A(b)) \right] & \mbox{for } s < s_M \\
  T_A(b) \, \exp \left(-\tilde{\sigma} T_A (b)/2 \right) & \mbox{for } s \geq s_M
  \end{array} \right.
\eeq
where the nuclear profile function, $T_A(b)$, is normalized to the atomic number, $A$,
and the shadowing correction, $S^{shad}(x_2,Q^2,b)$, for the gluons in the nucleus 
carrying a momentum fraction $x_2$ is taken from \cite{Tywoniuk:2007xy}. Note, that
no antishadowing effects are included in this model.

The "generalized" absorptive cross section that enters eq.~(\ref{eq:Sfe}) is a
function of $x_+$ \cite{Boreskov:1992ur,Arsene:2007gx}
\beq
\label{eq:sigmaabs}
\tilde{\sigma} (x_+) = \left[ (1 - \epsilon ) \Phi(t_{min}) + \epsilon x_+^\gamma 
\right] \sigma_\cc \;,
\eeq
where $\sigma_\cc$ is the total $\cc-N$ cross section, $\gamma =2 $ for charmonium
\cite{Boreskov:1992ur} and $\epsilon$ is 
a parameter 
characterizing the amount of longitudinal momentum lost in each rescattering.
The "form factor" 
\beq
\Phi(t_{min}) \;=\; \exp \left\{ - (x_2^c \big/ x_2)^2 \right\} \;,
\eeq
where $x_2^c \approx 0.1$ corresponds to the critical energy scale in 
eq.~(\ref{eq:sM}), controls the transition from the planar to the non-planar regime.
Note, that in the low-energy limit and at mid-rapidity, where no 
shadowing corrections exist, $S^{shad}=1$, the ordinary absorption in this model is given
by
\beq
\tilde{\sigma} \stackrel{M^2_\cc/s \sim 1}{\longrightarrow} \left[ (1- \epsilon) + 
\epsilon M^2_\cc / s\right] \sigma_\cc \;,
\eeq
whereas the Glauber result, without energy-momentum conservation, would be simply 
$\sigma_{abs} = (1-\epsilon)\sigma_\cc$. In the high-energy limit at mid-rapidity
, on the other hand,
\beq
\tilde{\sigma} \stackrel{s \gg M^2_\cc}{\longrightarrow} 0 \;,
\eeq
so that $S^{FE}=1$ and the suppression is fully governed by the shadowing corrections.
The parameters $\sigma_\cc$ and $\epsilon$ have been extracted from low energy data
in \cite{Arsene:2007gx}.

\begin{figure}
  \centering
  \includegraphics[width=0.8\linewidth]{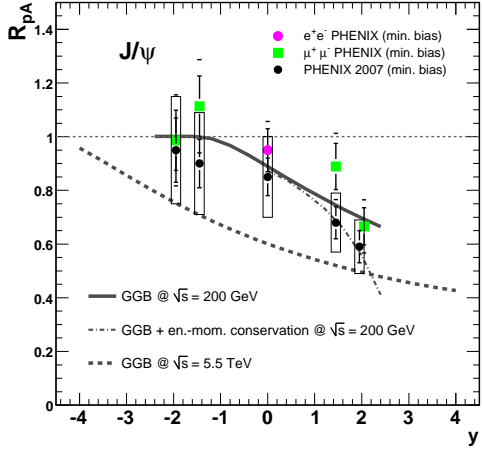}
  \caption{\label{fig:RapDepJpsi} Rapidity dependence of $J/\psi$ suppression in $pA$ 
  collisions at RHIC and LHC. Data are from \cite{Adler:2005ph,Adare:2007gn}.}
\end{figure}
In figure~\ref{fig:RapDepJpsi} we compare our calculations of gluon shadowing 
\cite{Tywoniuk:2007xy} (solid curve) and additionally with energy-mom-entum conservation
(dash-dotted) curve to data on $J/\psi$ in d+Au collisions at $\sqrt{s} = 200$ GeV 
\cite{Adler:2005ph,Adare:2007gn}. The latter effect becomes important already at 
$y = 2$. Note, that the same mechanism will also affect the forward yields of
charged hadrons at even smaller rapidities.

This constitutes the baseline in the search for 
the origin of anomalous suppression in high-energy $AA$ collisions. Also,
we also plot the predicted $J/\psi$ suppression in p+Pb collisions at LHC 
(see figure~\ref{fig:RapDepJpsi}). 
 
\section{Comover absorption and recombination of charmonium}
In the following section we will briefly describe the so-called co-mover
model for the case of charmonium secondary interactions. It was originally 
formulated to explain the anomalous $J/\psi$ suppression in $AA$ collisions at
the SPS \cite{Capella:1997,Armesto:1999,Capella:2000,Capella:2003}. Recently,
predictions were made for RHIC including only comover dissociation 
\cite{Capella:2005}.

As previously discussed, the possibility of recombination of $\cc$ pairs in the 
dense medium cannot be neglected at RHIC, due to the considerable density of open
charm at mid-rapidity. This recombination mechanism was first introduced in 
\cite{Thews:2000} neglecting spatial dependencies in the $J/\psi$ and open charm 
densities. The approach has recently been improved in 
\cite{Grandchamp:2004,Zhao:2007} and charmonium recombination has also been 
considered in statistical
hadronization models with charm conservation 
\cite{Braun-Munzinger:2000,Gorenstein:2001}. The dynamics of charmonium recombination
and dissociation has also been implemented in various scenarios in Monte-Carlo models
(see e.g. \cite{Linnyk:2007}).

The recombination mechanism has recently also been included in the co-movers 
interaction model (CIM) \cite{Capella:2007}.
Assuming a pure longitudinal expansion and boost invariance of the system, 
the rate equation which includes both dissociation and recombination effects 
for the density of charmonium at a given production point at impact parameter $s$ 
reads
\begin{align}
\label{eq:rateeq}
\tau \frac{d N_{J/\psi} (b,s,y)}{d \tau} \;=\; -\sigma_{co} \Big[ & N^{co}(b,s,y)
  N_{J/\psi}(b,s,y) \\ 
  & - \, N_c (b,s,y) N_{\bar{c}}(b,s,y) \Big] \;, \nonumber
\end{align}
where $N^{co}$, $N_{J/\psi}$ and $N_{c (\bar{c})}$ is the density of comovers, 
$J/\psi$ and open charm, respectively, and $\sigma_{co}$ is the interaction cross 
section for both dissociation of
charmonium with co-movers and regeneration of $J/\psi$ from $\cc$ pairs in the system
averaged over the momentum distribution of the participants. It is the constant of 
proportionality for both the dissociation and recombination terms due to detailed 
balance.

\begin{figure}[t!]
  \centering
  \includegraphics[width=0.8\linewidth]{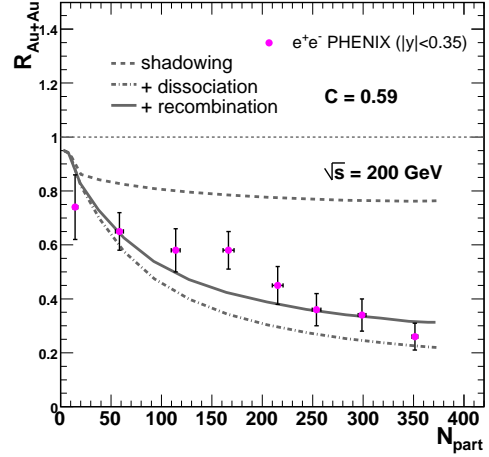} \\
  \includegraphics[width=0.8\linewidth]{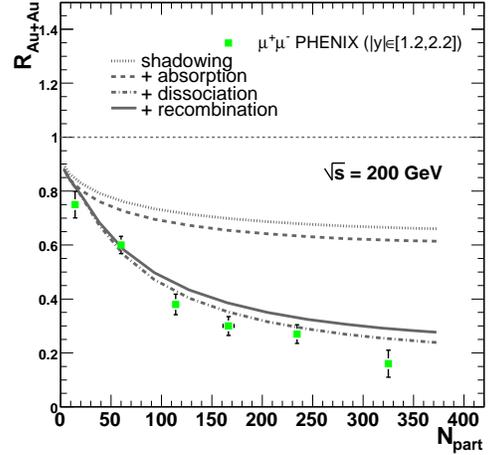}
  \caption{\label{fig:AuAuresult} Results for $J/\psi$ suppression in Au+Au
  collisions at $\sqrt{s} = 200$ GeV
  at mid- (upper figure) and forward (lower figure) rapidities. Data are taken from
  \cite{Adare:2007}.}
\end{figure}
Note, that the quantities that enter the rate equation, eq.~(\ref{eq:rateeq}), 
are densities of co-moving
matter at the same impact parameter as the $J/\psi$ produced in the initial hard scattering. 
This is quite different
from other approaches, e.g. 
\cite{Thews:2000} and also the statistical hadronization models 
\cite{Braun-Munzinger:2000,Gorenstein:2001}, where $c\bar{c}$ pairs in an 
extensive volume are allowed to recombine. In our case, the system is driven to a 
local equilibrium, given by
\beq
N_{J/\psi}(b,s,y) \,=\, \frac{N_{c}(b,s,y) \,N_{\bar{c}}(b,s,y)}{N^{co}(b,s,y)} \;.
\eeq

Equation~(\ref{eq:rateeq}) cannot be solved analytically. The suppression factor, 
i.e. the density of ratio of $J/\psi$ after the full evolution of the medium to the
one at some formation time $\tau_0$, can be approximated by
\begin{align}
S^{co}(b,s,y) = \exp \Big\{& - \sigma_{co} \big[ N^{co}(b,s,y) \nonumber \\
  & - C(y) N_{bin}(b,s) S^{shad}(b,s,y) \big] \\ & \times \, \ln \big[
  \frac{N^{co}}{N_{pp}(0)} \big] \Big\} \;,\nonumber
º\end{align}
where
\beq
\label{eq:C}
C(y) \;=\; \frac{\left( d \sigma^\cc_{pp} \big/ dy \right)^2}{\sigma^{ND}_{pp} d \sigma^{J/\psi}_{pp} \big/ dy} \;.
\eeq
Details of the model can be found in \cite{Capella:2007}. The quantities in 
eq.~(\ref{eq:C}) are all related to $pp$ collisions at the corresponding energy
and are taken from experiment. Formulated as above, the extension of 
CIM with inclusion of recombination effects does not involve any additional parameters.

With $\sigma_{co} = 0.65$ mb \cite{Armesto:1999,Capella:2000,Capella:2003} fixed 
from experiments at lower energies, we have calculated the suppression of $J/\psi$ in
Cu+Cu and Au+Au collisions at $\sqrt{s} = 200$ GeV in \cite{Capella:2007}. We present 
results for the Au+Au case for both mid- and forward rapidities in 
fig.~\ref{fig:AuAuresult}. Both the centrality and rapidity dependence are in good
agreement with the data.

The latter is due to a combination of stronger initial state effects at forward 
rapidities and the fact that recombination effects are weaker than at 
mid-rapidity. In a natural way, the CIM explains both of the observed $J/\psi$ 
puzzles in $AA$ collisions discussed earlier.

\section{$J/\psi$ suppression at LHC}
\begin{figure}[t!]
  \centering
  \includegraphics[width=0.8\linewidth]{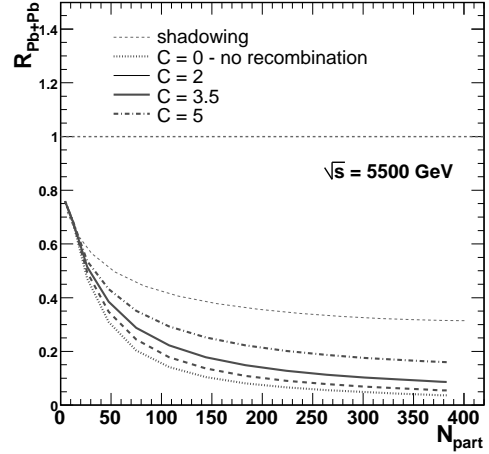} 
  \caption{\label{fig:predictionLHC} Prediction for $J/\psi$ suppression in 
  Pb+Pb collisions at mid-rapidity at LHC for different values of the parameter $C$.
  The upper line is the suppression due to gluon shadowing.}
\end{figure}
Predictions for LHC energy of $\sqrt{s} = 5.5$ TeV can be readily done assuming
that $\left. d \sigma_\cc \big/ d y \right|_{y=0} \sim 1 $ mb and $\sigma_{pp}^{ND} = 59$ 
mb which corresponds to $C = 2.5$; $\sigma_{co}$ is kept constant. They are presented 
in fig.~\ref{fig:predictionLHC}.

These predictions are in stark contrast with predictions of the statistical hadronization
model \cite{Braun-Munzinger:2008}, which favours a strong enhancement of $J/\psi$ 
for the most central collisions.

The reason of the discreapancy is the local form of our rate equation: only
comovers and open charm produced at the same impact parameter as the initial 
$J/\psi$ are allowed to interact. On the other hand, models assuming global
equilibrium of the produced charm with the medium allow for recombination of 
$\cc$ pairs from the whole volume of the fireball. A recent analysis suggests that
the thermal relaxation times of charmed quarks are rather long at RHIC, $\tau_c
\sim 5-7$ fm/c \cite{vanHees:2008}, but the situation is poorly known at LHC. 
Another point of dispute is the importance of initial state effects, which amount 
to almost 50$\%$ of the suppression in the CIM.

The 
large lever arm in energy and denisities of produced  will 
hopefully help us to understand and disentagle these issues in the future.

\section{Suppression of bottonium at LHC}
\begin{figure}[t]
  \centering
  \includegraphics[width=0.8\linewidth]{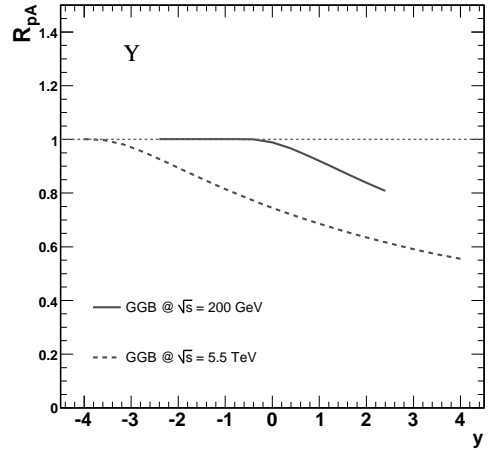}
  \caption{\label{fig:RapDepUps} Suppression of $\Upsilon$ at RHIC and LHC.}
\end{figure}

\begin{figure}[t]
  \centering
  \includegraphics[width=0.8\linewidth]{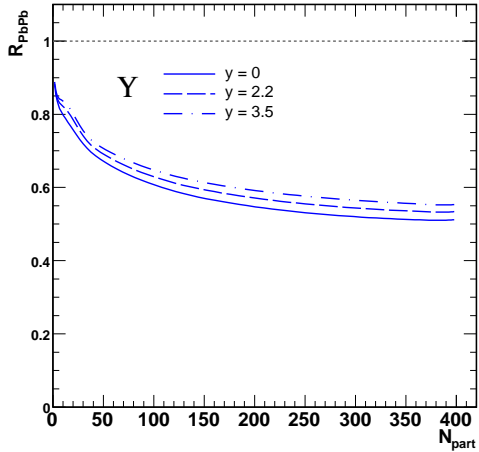}
  \caption{\label{fig:NpartUps} Centrality dependence of $\Upsilon$ suppression due
  to initial state effects (gluon shadowing) in Pb+Pb collisions at
  $\sqrt{s} = $ 5.5 TeV.}
\end{figure}
Finally, we present some results for bottonium at RHIC and at LHC, where excellent
detector capabilities will allow us to look in detail at the $\Upsilon$ family.
Since the $\Upsilon$ state is much smaller than charmonium, it may provide a clearer
probe of a thermalized system due to its presumably weaker interaction with matter.
On the other hand, due to strong binding energy, it may not dissolve at 
temperatures typical for heavy-ion collisions \cite{Karsch:1991}. Predictions
for LHC has previously been presented in \cite{Gunion:1997}.

Here, we would only like to discuss the impact of initial state effects on bottonium
production (for a recent discussion of final state medium effects we refer to
\cite{Grandchamp:2006}). The absorptive cross section for $\Upsilon$ is 40-50\% smaller
than the corresponding cross section for $J/\psi$ and $\psi'$. Furthermore, 
energy-momentum conservation mechanisms are pushed to higher $x_F$ due to the large
mass of the bottonium (corresponding to $\gamma \sim 3$ in eq.~(\ref{eq:sigmaabs})).
We expect therefore negligible nuclear absorption for RHIC and LHC kinematics.

Shadowing for bottonium in $pA$ collisions is shown in fig.~\ref{fig:RapDepUps} as 
a function of rapidity
at RHIC and LHC energies. Since our model \cite{Tywoniuk:2007xy} does not contain
antishadowing, the predictions for backward and mid-rapidities are uncertain 
up to the 10\% 
level, but we expect shadowing at forward rapidity at RHIC. At LHC, a large 
suppression is predicted for p+Pb collisions in most of the kinematics.

The suppression of bottonium due to gluon shadowing in Pb+Pb collisions at LHC is 
shown in fig.~\ref{fig:NpartUps} for several rapidities. The suppression is about
50\% from mid-central to central collisions, and would be the same for all
members of the $\Upsilon$ family. This establishes the baseline for further calculations
of bottonium dissociation and recombination in the final state.

%

\begin{acknowledgement}
K.T. would like to thank the organizers of HP2008 for financial support and 
acknowledges constructive discussions with K.~Boreskov, C.~Pajares and N.~Armesto.
This work was supported by the Norwegian Research Council (NFR) under contract No. 
166727 /V30, RFBF-6-2-17912, RFBF-06-02-72041-MNTI, INTAS 05-103- 7515, grant of 
leading scientific schools 845.2006.2, Federal Agency on Atomic Energy of Russia and
Program Ram\'on y Cajal of Spain.
\end{acknowledgement}


\vfill\eject
\end{document}